\documentclass{cimento}
\usepackage{amsmath}
\usepackage{longtable}
\usepackage{epsf,epsfig}
\usepackage{graphics,graphicx}
\usepackage{adjustbox}
\usepackage{braket}

\title{New theory highlights on $B_c$ decays}
\author{ N.~Losacco\from{ins:x}\from{ins:y}\thanks{Speaker}
        }
\instlist{\inst{ins:x} INFN, Sezione di Bari - via Orabona 4, 70125 Bari, Italy
  \inst{ins:y} Dipartimento di Fisica  ``M. Merlin'' Universit\`a e Politecnico di Bari - via Orabona 4, 70125 Bari, Italy}

\begin{document}

\maketitle

\begin{abstract}
We present recent results in semileptonic and non-leptonic exclusive $B_c$ decays to charmonium states both in $S$-wave, $J/\psi$ and $\eta_c$, and in $P$-wave, $\chi_{cJ}$ and $h_c$. The analysis, based on the heavy quark spin symmetry (HQSS), produces relations among form factors that parametrize the hadronic matrix elements in the amplitudes of the decays. These relations are helpful to control the hadronic uncertainty affecting these processes. Furthermore, $B_c$ decays allow us to get hint on the structure of states like $\chi_{c1}(3872)$, whose exotic or ordinary charmonium nature is debated.
\end{abstract}

\section{Introduction}

We focus on the heavy hadron decays induced by the $b \to c $ transition at the quark level. These decays allow us to measure a fundamental parameter of the Standard Model (SM), the element $|V_{cb}|$ of the Cabibbo-Kobayashi-Maskawa (CKM) mixing matrix. The analyses of several processes induced by the same quark transition but involving different hadrons in the initial or final states (exclusive/inclusive) must provide compatible results. This has not been achieved, yet, considering the $|V_{cb}|$ determinations from inclusive and exclusive $B$ decays which fuel discussions on  possible effects within the SM or beyond \cite{Gambino:2020jvv, CD1}. These decays provide also information on fundamental properties of the Standard Model, namely Lepton Flavour Universality (LFU). Signals of LFU violation have been detected in $B$ decays induced by this transition \cite{HFLAV:2022esi}. Such an observation would hint to physics beyond SM, the structure of which can be constrained by the analysis of various decay observables \cite{CD2,Alguero1,Cornella1}. 

Furthermore, these decays allow us to obtain information about the strong interaction between the quarks composing the mesons. For decays involving hadrons comprising a single heavy quark $Q$, a double  expansion in powers of $1/m_Q$ and of  $\alpha_s$ can be derived in QCD, providing a powerful method to classify the hadronic matrix elements, both for exclusive and  inclusive transitions \cite{BCD1, CDL1}. This is the basis for a control of the theoretical uncertainty in the measurements. For mesons comprising two heavy quarks such as $B_c$, the expansion parameter is the relative three-velocity of the heavy quarks, with counting rules  given by Non-relativistic QCD (NRQCD) \cite{Barmbilla1}.

\section{Semileptonic $B_c$ decays}

\begin{figure}[t]
\begin{center}
\includegraphics[width = 0.425\textwidth]{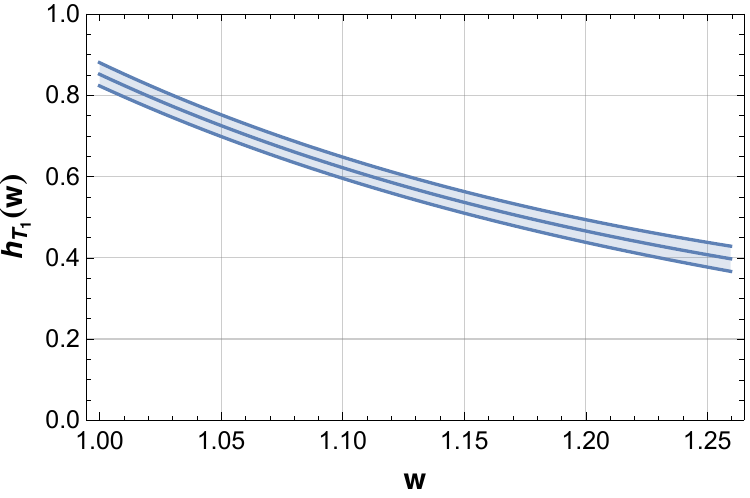}\hskip 0.2cm \includegraphics[width = 0.43\textwidth]{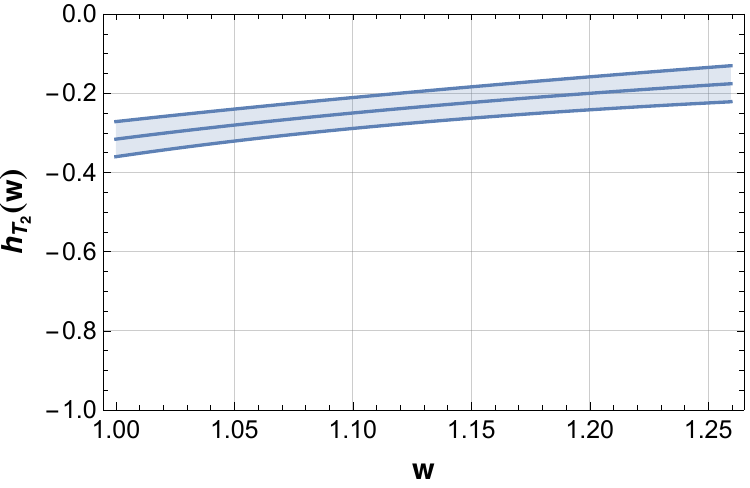}\\
\includegraphics[width = 0.43\textwidth]{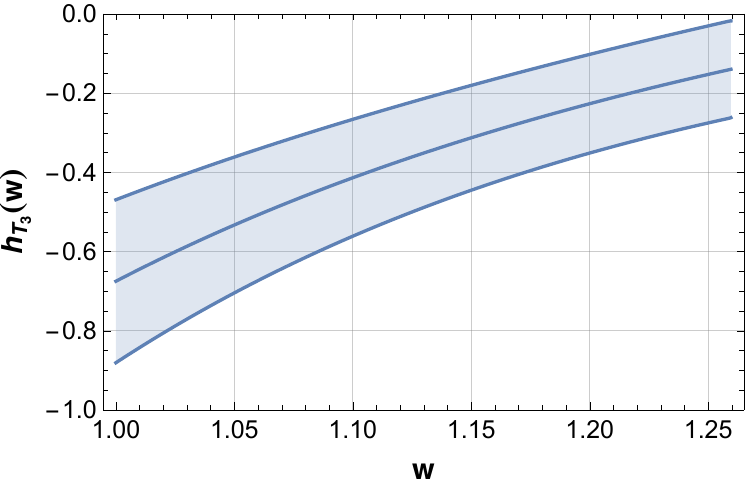}\hskip 0.45cm \includegraphics[width = 0.42\textwidth]{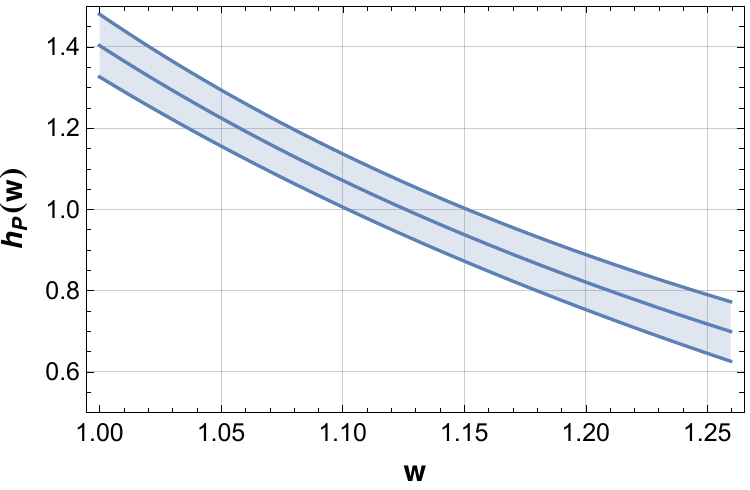}
    \caption{\small Tensor and pseudoscalar $B_c \to J/\psi$ form factors in the full kinematical range  and  using lattice QCD results as input.}\label{plotPsiFF}
\end{center}
\end{figure}

In \cite{Colangelo:2022lpy} relations have been obtained among the form factors governing $B_c \to J/\psi (\eta_c ) \ell \bar \nu_\ell$ near the zero-recoil point. Using the form factors $h_{A_1}$,$h_{A_2}$, $h_V$ determined by lattice QCD \cite{Harrison:2020gvo} and exploiting such relations, other form factors have been computed. As an example, fig.~\ref{plotPsiFF} shows the tensor and the pseudoscalar form factors for the transition to $J/\psi$ obtained from the relations in \cite{Colangelo:2022lpy} and the lattice QCD results \cite{Harrison:2020gvo}. The analysis has been also extended to the second order in $1/ m_Q$ \cite{Colangelo:2022lpy}.

%
%
%

\begin{figure}[t]
\begin{center}
\includegraphics[width = 0.45\textwidth]{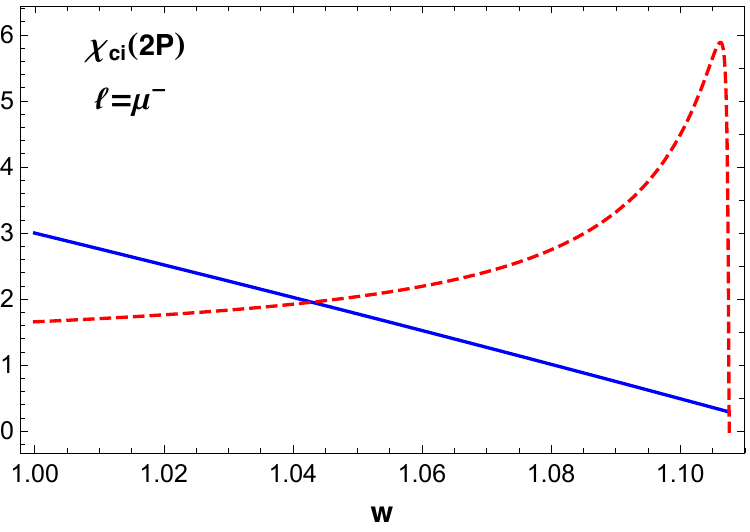}\hskip 0.4cm \includegraphics[width = 0.45\textwidth]{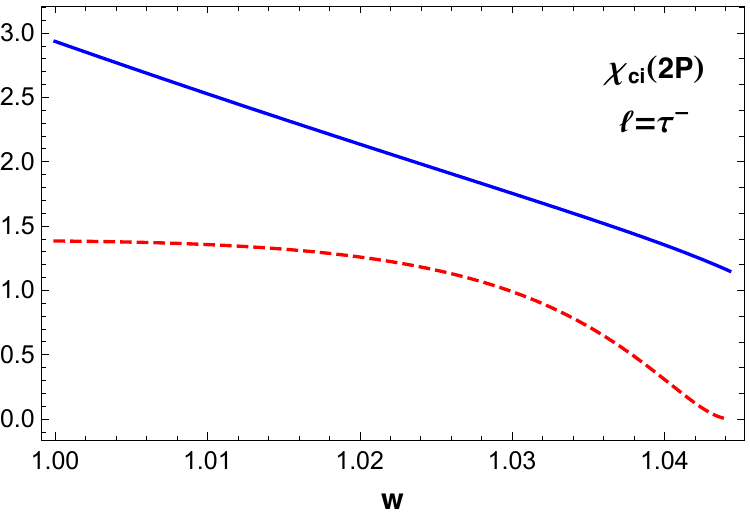}
    \caption{\small  Ratios of the decay distributions  $\frac{d \Gamma(B_c \to \chi_{c1} \ell \bar \nu)/dw}{d \Gamma(B_c \to \chi_{c0} \ell \bar \nu)/dw}$ (blue continuos line) and 
    $\frac{d \Gamma(B_c \to \chi_{c2} \ell \bar \nu)/dw}{d \Gamma(B_c \to \chi_{c1} \ell \bar \nu)/dw}$  (red dashed line) in the case $\ell=\mu$ (left) and $\ell=\tau$ (right) for the $2P$ final charmonia, using the LO relations among form factors extrapolated to the full kinematical range.}\label{fig1}
\end{center}
\end{figure}

The method has been applied to the analysis of the $B_c$ transitions to $P$-wave charmonia \cite{Colangelo:2022awx}. In a selected kinematical range the  $B_c$ and the $P$-wave charmonium matrix elements can be expressed as an expansion in the heavy quark relative three-velocity in the heavy hadrons, together with an expansion in the inverse heavy quark mass. In this range the form factors describing the decays of $B_c$ to the four $P$-wave charmonia (the lowest lying  or the  radial excitations)  are related. The relations among different modes can be experimentally verified. Their violation or confirmation can be used to obtain information on the nature of mesons belonging to the same spin multiplet. This is relevant for $\chi_{c1}(3872)$ (usually denoted as  $X(3872)$). This meson shows features hinting to a non conventional charmonium structure. Due to the closeness of its mass to the $D^{*0} \bar D^0$ threshold and  the large decay rate to  $J/\psi \, \pi \pi \, (J/\psi \, \rho )$  compared  to  $J/\psi \, \pi \pi \pi \, (J/\psi \, \omega)$, proposals have been put forward that $X(3872)$ is an exotic state of multiquark structure \cite{Brambilla2}. On the other hand, the  large  ratio  $\Gamma(X(3872)\to \gamma \, \psi(2S))/\Gamma(X(3872) \to \gamma \, J/\psi)$ and  the production cross sections in $e^+ e^-$ and $\gamma \gamma^*$  can be better accommodated identifying $\chi_{c1}(3872)$ with $\chi_{c1}(2P)$ \cite{Achasov:2022puf}. A confirmation of the relations obtained in \cite{Colangelo:2022awx} for $\chi_{c1}(3872)$ would point to the ordinary quark-antiquark structure. The behaviour expected for $\chi_{c1}(3872)$ considered as a radial excitation of $\chi_{c1}$ is shown in fig. \ref{fig1}.

\section{Non-leptonic $B_c$ decays}

The $B_c$ non-leptonic transitions to $P$-wave charmonia complement the information on the structure of states as $\chi_{c1}(3872)$ from the semileptonic modes. Using  the effective Hamiltonian describing the $b \to c \bar q_1 q_2$ transition with $q_{1,2}$ light quarks, we have evaluated the decay widths of  $B_c$ to $P$-wave charmonium and a light pseudoscalar/vector meson. The LO  relations for the form factors are used to predict several ratios of branching fractions.
The effective Hamiltonian governing the non-leptonic  $B_c$ decays to charmonium and a light meson is given by

\begin{equation}
\mathcal{H}_{eff}= \frac{G_F}{\sqrt{2}}V_{c b}^* V_{u q}\left(C_1 (\mu) Q_1 (\mu) + C_2 (\mu) Q_2 (\mu) \right) + {\text h.c.} \label{Hamil}
\end{equation}
\noindent where 
\begin{eqnarray}
Q_1 &=& \bar{u}_{\alpha} \gamma^{\mu} (1-\gamma_5) q_{\alpha} \bar{b}_{\beta} \gamma_{\mu} (1-\gamma_5) c_{\beta} \nonumber \\
Q_2 &=& \bar{u}_{\alpha} \gamma^{\mu} (1-\gamma_5) q_{\beta} \bar{b}_{\beta} \gamma_{\mu} (1-\gamma_5) c_{\alpha} \, .
\end{eqnarray}
$\alpha, \beta$ are colour indices, and 
$q$ identifies the down-type quark of the final light meson.  $\mu$ represents the scale dependence of the Wilson coefficients which encode the short-distance physics from energy greater than $\mu$. It is cancelled in the amplitude by the $\mu$ dependence of the operators matrix elements. In the computation of $B_c$ decays we set $\mu = m_b$. After a Fierz transformation and discharging the colour-octet operator, we obtain the Hamiltonian 
\begin{equation}
\mathcal{H}_{eff}= \frac{G_F}{\sqrt{2}} V_{c b}^* V_{u q} a_1 (\mu) Q_1 (\mu)  \, , \label{Hamilnew}
\end{equation}
where
\begin{equation}
a_1 = C_1 + \frac{1}{N_c} C_2.
\end{equation}
\noindent
For the processes  $B_c^+ \to M_{c \bar{c}}(P) \, M_{P (V)}$, where  $M_{c \bar{c}}(P)$ is one of the $P$-wave charmonium and  $P$,$V$  a light pseudoscalar or a vector meson, the decay width

\begin{eqnarray}
\Gamma (B_c^+ \to M_{c \bar{c}}(P) \, M_{P (V)}) = \frac{\lvert \textbf{q} \rvert}{8 \pi m_{B_c}^2} \lvert \mathcal{A}(B_c^+ \to M_{c \bar{c}}(P) \, M_{P (V)})\rvert ^2,
\label{eq:gamma}
\end{eqnarray}
with  $\lvert \textbf{q} \rvert = \frac{1}{2 m_{B_c}} \sqrt{\lambda(m_{B_c}^2,m_{M_{c \bar{c}}}^2,m_{P(V)}^2)} $ and  $\lambda(x,y,z)$ the K\"all\'en function, involves the matrix element 

\begin{eqnarray}
\mathcal{A}(B_c^+ \to M_{c \bar{c}}(P) \, M_{P (V)}) = \bra{M_{c\bar{c}}(P) \, M_{P(V)}} \mathcal{H}_{eff} \ket{B_c^+}.
\label{eq:ampl1}
\end{eqnarray}
The amplitude is factorized as 

\begin{eqnarray}
\mathcal{A}(B_c^+ \to M_{c \bar{c}}(P) \, M_{P (V)}) = \frac{G_F}{\sqrt{2}} V_{c b}^* V_{uq} a_1 (\mu) \bra{M_{c \bar{c}}(P)}\bar{b} \gamma_{\mu} (1-\gamma_5) c \ket{B_c^+} \bra{M_{P (V)}}\bar{u} \gamma^{\mu} (1-\gamma_5) q \ket{0}. \nonumber
\label{eq:ampl}
\end{eqnarray}
This involves the current-particle matrix elements
\begin{eqnarray}
\bra{M_P(q)}\bar{u} \gamma^{\mu} (1 &-& \gamma_5) q \ket{0} = - i f_{P} q^{\mu} \label{AnnMat}\\ \bra{M_V(q,\epsilon_V)}\bar{u} \gamma^{\mu} (1 &-& \gamma_5) q \ket{0} = m_V f_{V} \epsilon^{*\mu}_V  \, , \label{AnnMatVec} 
\end{eqnarray}
and $\bra{M_{c \bar{c}}(P)}\bar{b} \gamma_{\mu} (1-\gamma_5) c \ket{B_c^+}$ for which we used the parametrization in \cite{Colangelo:2022awx}.
The use of the naive factorization approach is based on the Bjorken's colour transparency argument \cite{BJColor}. It is possible to partially encode nonfactorizable effects replacing the coefficient  $a_1(\mu)$  ($a_2(\mu)$ for colour suppressed processes),  by the  effective coefficient $a_1^{eff}(\mu)$ ($a_2^{eff}(\mu)$) treated as a phenomenological parameter.

For the hadronic form factors of $B_c$ to $ \chi_{cJ}$ and $h_c$, we use the results described in the previous section at LO in $1/m_Q$ \cite{Colangelo:2022awx}. Remarkably, at LO some form factors vanish: a consequence is that the $B_c$  transition to $\chi_{c1}$  and a light pseudoscalar is suppressed.
\noindent

The results for the branching fractions for $\pi^+$ in the final state are in Table \ref{tab:piK}, together with the results for the $2P$ excitations and for the case with $K^+$ in the final state.

\begin{table}[!h]
\begin{adjustbox}{}
\begin{tabular}{c|ccc}

  & $\frac{\mathcal{B}(B_c^+ \to \chi_{c0} \, \pi^+)}{\mathcal{B}(B_c^+ \to \chi_{c2} \, \pi^+)}$  & $\frac{\mathcal{B}(B_c^+ \to h_c \, \pi^+)}{\mathcal{B}(B_c^+ \to \chi_{c0} \, \pi^+)}$ & $\frac{\mathcal{B}(B_c^+ \to h_c \, \pi^+)}{\mathcal{B}(B_c^+ \to \chi_{c2} \,\pi^+)}$ \\
\hline
$1P$ & $0.658$ & $2.429$ & $1.597$ \\
$2P$ & $0.583$ & $2.746$ & $1.601$  \\
\hline \hline
 & $\frac{\mathcal{B}(B_c^+ \to \chi_{c0} \, K^+)}{\mathcal{B}(B_c^+ \to \chi_{c2} \, K^+)}$  & $\frac{\mathcal{B}(B_c^+ \to h_c \, K^+)}{\mathcal{B}(B_c^+ \to \chi_{c0} \, K^+)}$ & $\frac{\mathcal{B}(B_c^+ \to h_c \, K^+)}{\mathcal{B}(B_c^+ \to \chi_{c2} \, K^+)}$ \\
\hline
$1P$ & $0.663$ & $2.482$ & $1.645$ \\
$2P$ & $0.586$ & $2.845$ & $1.668$  
\end{tabular}
\end{adjustbox}
\caption{\small Ratios of branching fractions of $B_c$ decays to charmonium state and $\pi^+$ or $K^+$ meson.}
\label{tab:piK}
\end{table}
\noindent

The modes with $\chi_{c1}(1P)$ and $\chi_{c1}(2P)$  are suppressed: the observation of such a suppression for $\chi_{c1}(3872)$ would favour the identification of $X(3872)$ as an ordinary charmonium state.
The same results are obtained for the $\rho^+$ and $K^{*+}$ modes \cite{Losacco}. The conclusion is that the production of $\chi_{c1}$ is suppressed and of $h_c$ is enhanced compared to the other charmonia.

\section{Conclusions}

We have investigated a method based on heavy quark spin symmetry and the use of NRQCD power counting to obtain a systematic treatment of the form factors for the $B_c$ to charmonium decays. The formalism allows us to organize the states in doublets and multiplets, obtaining relations between form factors and consequently among observables in different decay modes. Using lattice QCD results \cite{Harrison:2020gvo} for a few form factors entering in the description of the decay we can obtain others, namely the form factors parametrizing the matrix element of the tensor and pseudoscalar operator for the transition to $J/\psi$, and combinations of them in the case of $\eta_c$. The relations for the $P$-wave states can be employed to get information on $\chi_{c1}(3872)$. If this state behaves according to our predictions, there is a hint to the ordinary charmonium structure.

The non-leptonic decays of $B_c$ mesons have been computed using a naive factorization approach. We focused on the decays involving the lowest-lying and first radial excitations of the $P$-wave charmonium states together with $\pi^+$, $K^+$, $\rho^+$, and $K^{*+}$. By employing the leading-order (LO) relations among the form factors for $B_c \to \chi_{cJ}(h_c)$ transitions \cite{Colangelo:2022awx}, we have predicted several branching ratios. We have found that the $\chi_{c1}$ channel, both in the $1P$ and $2P$ multiplets, is suppressed. If $\chi_{c1}(3872)$ is a conventional charmonium we should observe such a suppression.

\acknowledgments
I thank Pietro Colangelo, Fulvia De Fazio, Francesco Loparco and Martin Novoa-Brunet for collaboration and useful discussions. The work is carried out within the INFN projects
(Iniziative Specifiche) QFT-HEP and SPIF.

\end{document}